\documentstyle{l-aa}

\begin{document}

   \thesaurus{03         % Extragalactic sources
              (11.01.2;  % Galaxies:active,
               11.03.2;  % Galaxies:compact,
               11.05.2;  % Galaxies:evolution,
               11.09.1 %0108+388;  % Galaxies:individual,
               11.10.1;  % Galaxies:jets,
               13.18.1)} % Radio continuum: galaxies.

   \title{Renewed Radio Activity of Age 370 years in the Extragalactic Source 0108+388.}

   \author{I.Owsianik 
          \inst{1}
          \and J.E.Conway
          \inst{2}
          \and A.G. Polatidis
          \inst{2}
          \inst{,3}
          }
            
  \offprints{I.Owsianik}

   \institute{Toru\'n Centre for Astronomy,
                Radio Astronomy Dep.,
                ul.Gagarina 11, PL 87 100 Toru\'n, Poland\\
                e-mail iza@astro.uni.torun.pl
              \and 
                Onsala Space Observatory,
                S 439 92 Onsala, Sweden
               \and
                Joint Institute for VLBI in Europe,
                Postbus 2, NL 77990, AA Dwingeloo, The Netherlands
                 }

   \date{Received month day, year; accepted month day, year}

   \maketitle
   
   \begin{abstract}

We present  the  results of  multi-epoch  global VLBI  observations  of the
Compact  Symmetric Object (CSO) 0108+388 at 5 GHz.  Analysis of data spread
over 12 years shows strong  evidence for an increase in the  separation  of
the  outer  components  at a rate  of  $0.197\pm0.026  h^{-1}c$.  Given  an
overall  size of 22.2  $h^{-1}$pc  this  implies  a  kinematic  age of only
$367\pm48$ yrs.  This result strongly supports the idea that radio emission
in Compact Symmetric Objects arises from recently  activated radio sources.
The  presence  of weak  radio  emission  on  kpc-scales in  0108+388
suggests  recurrent  activity in this source, and that we are  observing it
just as a new period of activity is beginning.

      \keywords{Radio Continuum: Galaxies
                --- Galaxies:active --- compact --- evolution --- 
                jets --- individual: 0108+388
               }
   \end{abstract}

%____________________________________________________

\section{Introduction}

There  exists  a  class  of  powerful  radio  sources  consisting  of  high
luminosity radio emission regions separated by less than 1 kpc and situated
symmetrically  about the  centre of  activity.  Objects  of this  type were
first    described   as   `Compact    Doubles'   by   Phillips   \&   Mutel
(\cite{phillips}),  but the more generic name of Compact  Symmetric Objects
(CSOs) was given by Wilkinson et al.  (\cite{wilkinson}).

Three  possible evolutionary scenarios of CSOs  have been proposed: 1)
they are  old  `frustrated'  sources,  in which a  dense   environment
doesn't allow them  to grow (van Breugel  et al.   \cite{breugel}); 2)
they are young sources which will `fizzle out'  after a short lifetime
(Readhead  et al.  \cite{read94})  or 3)  they  represent a very young
stage in  the evolution of  large-sized classical radio  sources (e.g.
Fanti  et al.\cite{fanti},  Readhead et  al.  \cite{read96b}, Begelman
\cite{begelman}, Owsianik \& Conway \cite{owsianik}). 

One archetypical CSO is the  radio source 0108+388.  This radio object
is identified with a   galaxy of  m$_{V}$=22.0  mag at  redshift  {\it
z}=0.669 (Lawrence et  al.  \cite{lawrence}).  The radio  flux density
is weakly polarised  ($0.30\%\pm0.08\%$ at 4.8  GHz) and does not show
significant variations (Aller et  al.  \cite{aller}).  0108+388  has a
spectral   turnover    around   5    GHz,  with    spectral    indices
$\alpha^{22\mathrm{GHz}}_{\mathrm{5GHz}}=~-1.27$  and
$\alpha^{5\mathrm{GHz}}_{0.6\mathrm{GHz}}=2.1$
($S~\propto~\nu^{~\alpha}$, Baum et al.  \cite {baum}).

\section {Observations and imaging}

There have been three epochs of global VLBI  observations  of 0108+388 at 5
GHz evenly  spread  over a period of 12 years.  The first two  epochs  were
made on 8th Dec 1982 with a global  array of 5  telescopes  and on 23rd Nov
1986  using   multiple   snapshots   with  9  telescopes   (Conway  et  al.
\cite{con94}).  Here we  reanalyse  these  epochs  and add new data  from a
multi-snapshot  10 station global VLBI observation  made on 18th Sep 1994.
The  telescopes  used included  those from the European VLBI Network (EVN),
the Very  Long  Baseline  Array  (VLBA),  the Very  Large  Array  (VLA) and
Haystack Observatory.

\begin{figure}[htbp]
        \vspace{5.5cm}
        \includegraphics{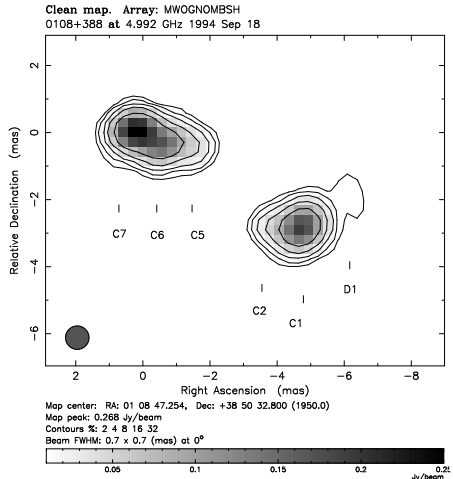}
        \vspace{4mm}
\caption{{\footnotesize Third  epoch image of 0108+388 with 
the positions of components identified in gaussian modelfitting
indicated. Rms noise=0.88 mJy beam$^{-1}$.}}
\label{Fig.1}
\end{figure}

The amplitude  calibration,  fringe-fitting,  imaging and modelfitting were
performed  following  the  procedures   described  in  Owsianik  \&  Conway
(\cite{owsianik}).  Figure~\ref{Fig.1}  shows  the  highest  dynamic  range
image  obtained  from the third epoch data using  DIFMAP  (Shepherd  et al.
\cite{shepherd95}),  which was used as a starting  point in  remapping  the
other two epochs.  Modelfitting  of gaussian  components to the  visibility
data was also  carried  out at each  epoch  using the 3rd epoch  model as a
starting  model (see  Table~\ref{Tab.1}).  In our  modelfitting  we allowed
only the flux densities and positions of components to vary.  The estimated
models  showed  a good  fit to  the  visibility  data  with  total  reduced
$\chi^{2}$  agreement  factors  of  Q$_{TOT}$=1.158,   Q$_{TOT}$=1.255  and
Q$_{TOT}$=1.082 for epochs 1 to 3 respectively.

%___________________________Fig.1_______________________________________

%________________________________________________________________

\section {Results}

\subsection {Multi-Epoch Intercomparison}

The image of the third epoch data of 0108+388  (Fig.~\ref{Fig.1}) shows the
pc-scale  structure  of the source at 5 GHz, which  consists  of two bright
regions with embedded  compact  components  at the leading  edges, which we
interpret as hotspots.  The images and modelfits are consistent  with those
of Taylor et al.  (\cite{taylor})  at 15 GHz, whose  nomenclature we use in
this  paper.  In  addition  at 15 GHz Taylor et al.  (\cite{taylor})  found
evidence for a Synchrotron Self-Absorbed (SSA) core region connected to the
outer regions by jet-like  structures.  Due to its inverted  spectrum we do
not see this core component in our 5 GHz image.

\begin{figure}[htbp]
        \vspace{3.5cm}
        \includegraphics{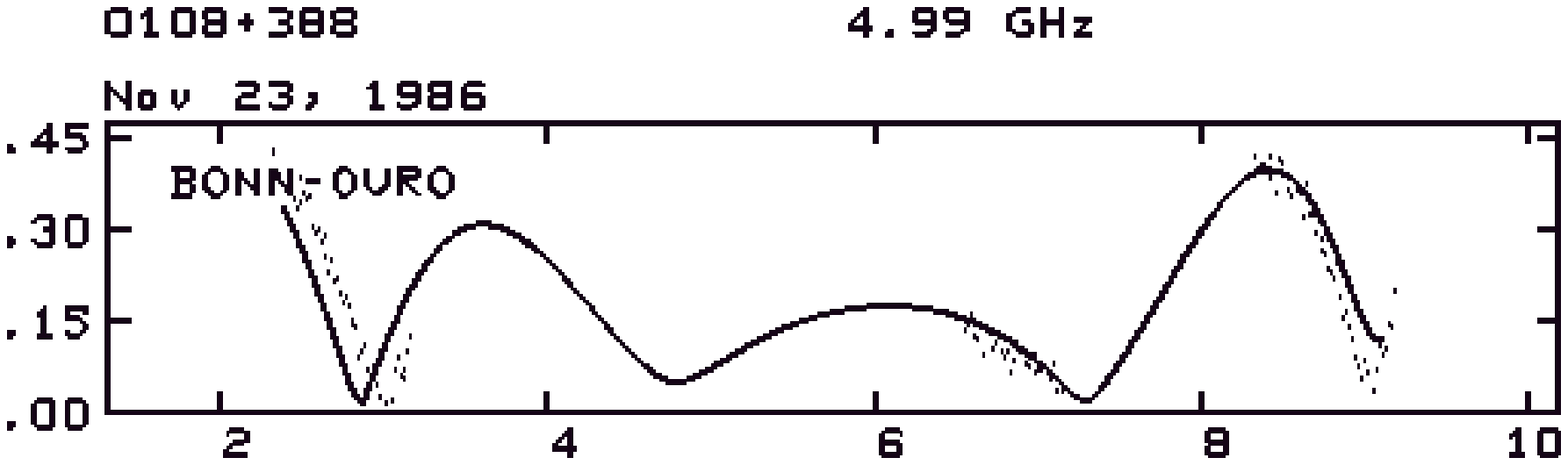}
        \includegraphics{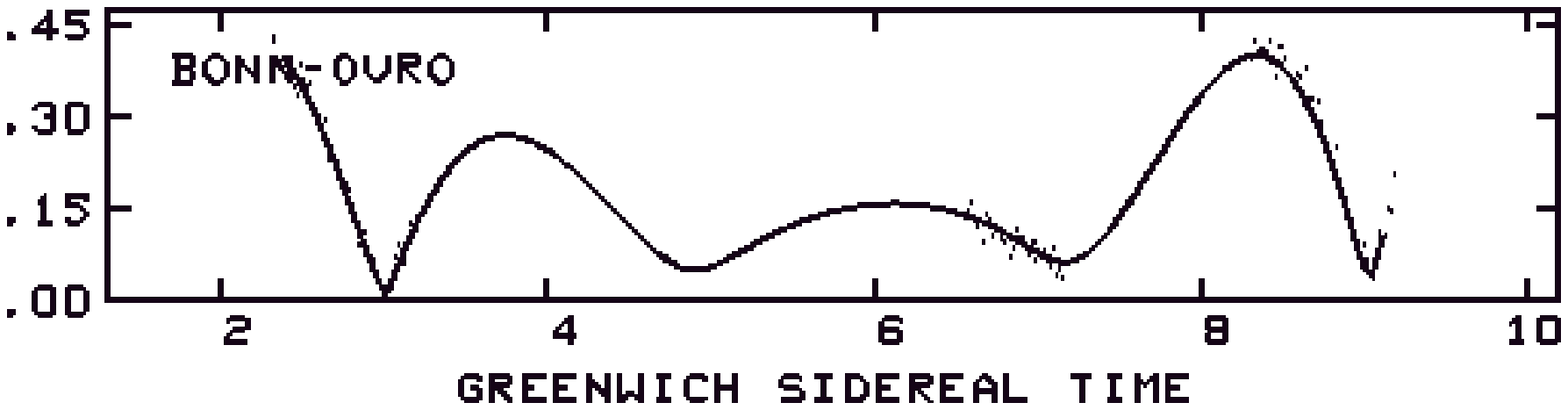}
        \includegraphics{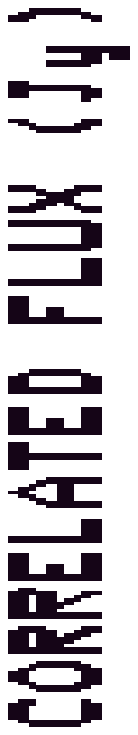}
\caption{{\footnotesize Temporal changes in visibility data from 0108+388
at 5 GHz. {\it Top}:
Third epoch model plotted against the second epoch data, {\it Bottom}:
the same second epoch data with the best fitting model from gaussian modelfitting 
after allowing component positions to change.
}}
\label{Fig.2}
\end{figure}

Examining the visibility  data we see clear evidence on the long  baselines
for   structural   changes   in   0108+388   between   epochs   2   and   3
(Fig.~\ref{Fig.2})  which are explained by changes in component  positions.
Consistent  differences are found on comparing the data from epochs 1 and 2
(Conway  et al.  \cite{con94}).  Linear  regression  fits  to the  observed
changes in the gaussian  component  separation  between the three epochs of
data give us estimates of relative component motions and associated errors.
From this analysis we estimate an angular  separation rate of components C1
and C7 of  $9.27\pm1.21  \mu$as  yr$^{-1}$  corresponding  to a velocity of
$0.196\pm0.026    h^{-1}c$   (for    $q_{0}$=0.5    and   $H_{0}$=100   $h$
kms$^{-1}$Mpc$^{-1}$).  This result is  consistent  with earlier  estimates
based on two epoch  data sets and  shorter  time  baselines.  Conway et al.
(\cite{con94})  found a velocity of 0.18 $h^{-1}c$ based on the first two 5
GHz epochs, which was cautiously claimed only as an upper limit.  Taylor et
al.  (\cite{taylor}) also estimated  $0.22\pm0.20  h^{-1}c$ by comparing 10
GHz  and 15 GHz  observations  from  1984  and  1994.

%_____________________________Table.1_________________________________
\begin{table}[htbp]
      \caption[]{Gaussian model for the 3rd epoch of 0108+388}
         \label{Tab.1}
         \begin{flushleft}
         \begin{tabular}{lllllll}
            \hline\noalign{\smallskip}
{\small Comp.}& {\small S}& {\small r}& {\small $\Theta$}& {\small a}&  {\small $b/a$}& {\small $\Phi$}\\
& ({\small Jy})& ({\small mas})& ({\small $\degr$})& ({\small mas})& & ({\small $\degr$})\\
            \noalign{\smallskip}
            \hline
            %\noalign{\smallskip}
C7...& 0.490& 0.19&   78.98& 0.78& 0.61& -87.75\\         
C6...& 0.253& 0.78& -120.62& 0.82& 0.48& -62.53\\  
C5...& 0.063& 1.64& -106.95& 0.59& 0.78& -16.34\\
        \hline\noalign{\smallskip}
C2...& 0.045& 4.72& -126.36& 0.38& 0.51& -90.00\\         
C1...& 0.422& 5.51& -120.76& 0.85& 0.78& -45.52\\  
D1...& 0.030& 6.52& -107.21& 1.10& 0.38&  12.84\\
         %\noalign{\smallskip}
         \hline
        \end{tabular}
        \end{flushleft}
\begin{list}{}{}
\item[] 
{\small Parameters of the Gaussian components: S---flux density;
r,$\Theta$---polar coordinates, with polar angle measured from the North
through East;
a,b---major and minor axes of the FWHM contour;
$\Phi$---position angle of the major axis.}
\end{list}
\end{table}

%_________________________________________________________________

We also found from our three epoch data that the C6-C7 separation increased
at  $0.088\pm0.018   h^{-1}c$.  Finally  we  see  evidence  for  motion  in
component  C5,  which is  moving  towards  component  C7 at a  velocity  of
$0.942\pm0.151  h^{-1}c$.  A motion of C5  relative to C7 with  velocity of
$0.92\pm0.20  h^{-1}c$ was also detected by Taylor et al.  (\cite{taylor}).

\subsection{Source orientation and true velocity of the components}

Since at 5 GHz we were  unable to  detect  the core  component,  we  cannot
measure the individual motions of C1 and C7 relative to the core.  However,
we can constrain the hotspot advance speeds and the overall  orientation of
the source using the observed total  separation  rate between C1 and C7 and
the  `arm-length   ratio'  of  1.3  measured  at  15  GHz  (Taylor  et  al.
\cite{taylor}).  Assuming  both  hotspots   have  the  same  advance  speed
$\beta_{hot}$ through the surrounding medium and assuming this velocity has
remained  constant over the lifetime of the source then the apparent source
asymmetry  must be due to  light  travel  time  effects.  Given  this,  the
measured     `arm-length    ratio'    provides    the    constraint    that
$\beta_{hot}\cos\theta=0.13$,  where  $\theta$ is the angle between the jet
axis and the line of sight.  The apparent  separation velocity $v_{sep}$ of
C1 and C7 is connected to $\beta_{hot}$ and $\theta$ by the relation:
\begin{equation}        
v_{sep}=\frac{2c\beta_{hot}\sin\theta}{1-\beta_{hot}^{2}\cos^{2}\theta}\,,  
\end{equation} 
where  $v_{sep}$  depends on  $H_{0}$  and  $q_{0}$  as well as the
measured angular rate of separation.  From our two constraints, for a
given choice of $h$ and $q_{0}$ we can solve uniquely for both $\theta$
and    $\beta_{hot}$.   For    $q_{0}=0.5$    and   $h=0.5$   we   find
$\theta=56^{\circ}\pm11^{\circ}$    and    $\beta_{hot}=0.23\pm0.02$,   for
$q_{0}$=0.5  and  $h=1.0$  then  $\theta=36.6^{\circ}\pm6^{\circ}$  and
$\beta_{hot}=0.16\pm0.01$.

For the jet component C5, given its apparent  motion relative to C7 and the
`arm-length  ratio', we can estimate its  apparent  motion  relative to the
core.  If we assume that C5 has the same  $\theta$ as the hotspots then for
a given $q_{0}$ and $h$ we can derive the true  velocity  $\beta_{jet}$  of
C5.  Alternatively  we can assume that  $\beta_{jet}<1$  and constrain  the
allowed  cosmological  parameters.  For $q_{0}=0.5$, we find solutions with
$\beta_{jet}<1$   only  for  values  of  Hubble   constant   greater   than
$H_{0}=54\pm6$  kms$^{-1}$Mpc$^{-1}$  while for  $q_{0}=0$  a larger  lower
limit  to the  Hubble  constant  ($H_{0}=80\pm8$  kms$^{-1}$Mpc$^{-1}$)  is
required.   In   most   jets   it   appears   that   $\gamma>2$   so   that
$\beta_{jet}>0.9$  (Taylor  \&  Vermeulen  \cite{tay97}).  Therefore  for a
given  $q_{0}$ the best  estimate  of $H_{0}$ is close to the lower  limits
given above (e.g.  $H_{0}=54$ kms$^{-1}$Mpc$^{-1}$ for $q_{0}=0.5$).

%___________________________Fig.2_______________________________________

%________________________________________________________________

\subsection{Physical parameters of the source}

Dividing the distance  between C1 and C7 by their observed  separation rate
we estimate  that the pc-scale  source would have had zero size  $367\pm48$
yrs ago  (measured in the source  frame).  This  implies  that the pc-scale
structure  of  0108+388  is very  young.  Furthermore,  from  the  measured
angular  expansion  rate and  using  the  arguments  given in Sect.  3.2 we
estimate a minimum  physical  advance  speed of the hotspots in 0108+388 of
$0.16c$.  Combining this with our  estimation of the internal  pressures of
the  hotspots   ($1.14\times10^{-4}$dyne   cm$^{-2}$)  from   equipartition
arguments and given ram pressure  confinement  of the hotspots we derive an
upper limit on the external density of 2.1 cm$^{-3}$.

Given our  estimates  of the age and the jet thrust  (found by dividing the
internal  pressure  in  each  hotspot  by  its  area)  we can  compare  the
mechanical luminosity required to drive the hotspots forward with the radio
luminosity  and jet power.  For an age of 367 yrs the  combined  mechanical
luminosity   of  the  two   hotspots   is   $1.17\times10^{44}   h^{-17/7}$
ergs$^{-1}$,  while the observed  radio  luminosity  of the two hotspots is
about $7.76\times10^{43} h^{-2}$ ergs$^{-1}$.  Following the arguments used
in  Readhead et al.  (\cite{read96a})  the upper  limit on the total  power
supplied by the jets is $1.37\times10^{45}h^{-10/7}$ ergs$^{-1}$.  A lower
limit on the total jet power can be obtained by adding  together  the radio
power and mechanical work.  The total jet luminosity is (for $h=0.54$) then
in  the  range   $7.89\times10^{44}$   ergs$^{-1}$  to  $3.29\times10^{45}$
ergs$^{-1}$  and the  efficiency  of conversion  of the jet energy to radio
emission is between 8\% and 34\%.  In contrast,  for  classical  FRII radio
galaxies upper limits on the hotspot radiative efficiencies are only of the
order of a few percent (Owsianik \& Conway \cite{owsianik}).

\section{Discussion}

From three epoch  observations  we find strong  evidence for an increase in
separation of the two outer pc-scale  components of 0108+388 (C1 and C7) at
a velocity of  $0.197~\pm~0.026  h^{-1}c$.  These outer components have all
the  properties  expected of hotspots in a CSO (in contrast for instance to
being knots in a two sided jet):  the two components lie at the extremities
of the pc-scale  structure,  have simple SSA spectra and are connected  via
jets to a core (Taylor et al.  \cite{taylor}).  Given their  identification
as hotspots the measured  separation  rate of components  C1 and C7 implies
that the pc-scale source in 0108+388 is very young, i.e.  $367\pm48$ yrs.

Recent  results from other CSOs suggest that they are also very young radio
sources (e.g.  Fanti et al.  \cite{fanti}, Readhead et al.  \cite{read96b},
Owsianik \& Conway  \cite{owsianik}).  The evidence therefore suggests that
CSOs are not in general  `frustrated' slowly growing sources within a dense
environment  but are instead fairly rapidly ($\sim 0.2c$) growing  sources.
There are then several  possibilities for the subsequent evolution of CSOs.
The simplest  possibility is that they evolve via a Compact Steep  Spectrum
phase into  classical  double radio  sources  (FRIs or FRIIs).  Given their
fairly rapid  expansion  rate the sources would only be expected to spend a
short time in their CSO phase.  In order to explain the large  fraction  of
CSOs in flux limited  samples  there must be a strong  negative  luminosity
evolution   with   increasing   source   size   (e.g.   Readhead   et   al.
\cite{read96b}).  If  such  luminosity  evolution  occurs  the  `population
problem' for CSOs is explained  because  CSOs then evolve into much weaker,
more numerous sources.  Such negative  luminosity  evolution is expected in
theory (e.g.  Begelman  \cite{begelman}),  and evolution of just the amount
required is found if we compare the efficiency of radio  production in CSOs
and classical  radio double sources (e.g.  Readhead et al.  \cite{read96a},
Owsianik \& Conway \cite{owsianik}).

Another  way to  explain  the  large  population  of CSOs is via  recurrent
activity in these sources  (e.g.  O'Dea \& Baum  \cite{dea97},  Reynolds \&
Begelman  \cite{reynolds}).  In  this  model  the  sources  are  quite  old
but the activity  occurs in short  bursts.
Every time the activity  restarts the jet must propagate  again from
the  nucleus  to the kpc  working  surface.  Hence  the  sources  appear as
Compact  Symmetric  Objects (size $< 1$ kpc) for a significant  fraction of
their  lifetime.  The source  0108+388  appears to be a good  candidate for
such  recurrent  activity,  since in this source Baum et al.  (\cite{baum})
have detected weak extended emission $\sim20 \arcsec$ ($\sim78 h^{-1}$ kpc)
to the East of the nucleus.

%___________________________Fig.3_______________________________________

\begin{figure}[htbp]
    \vspace{5.5cm}       
\includegraphics{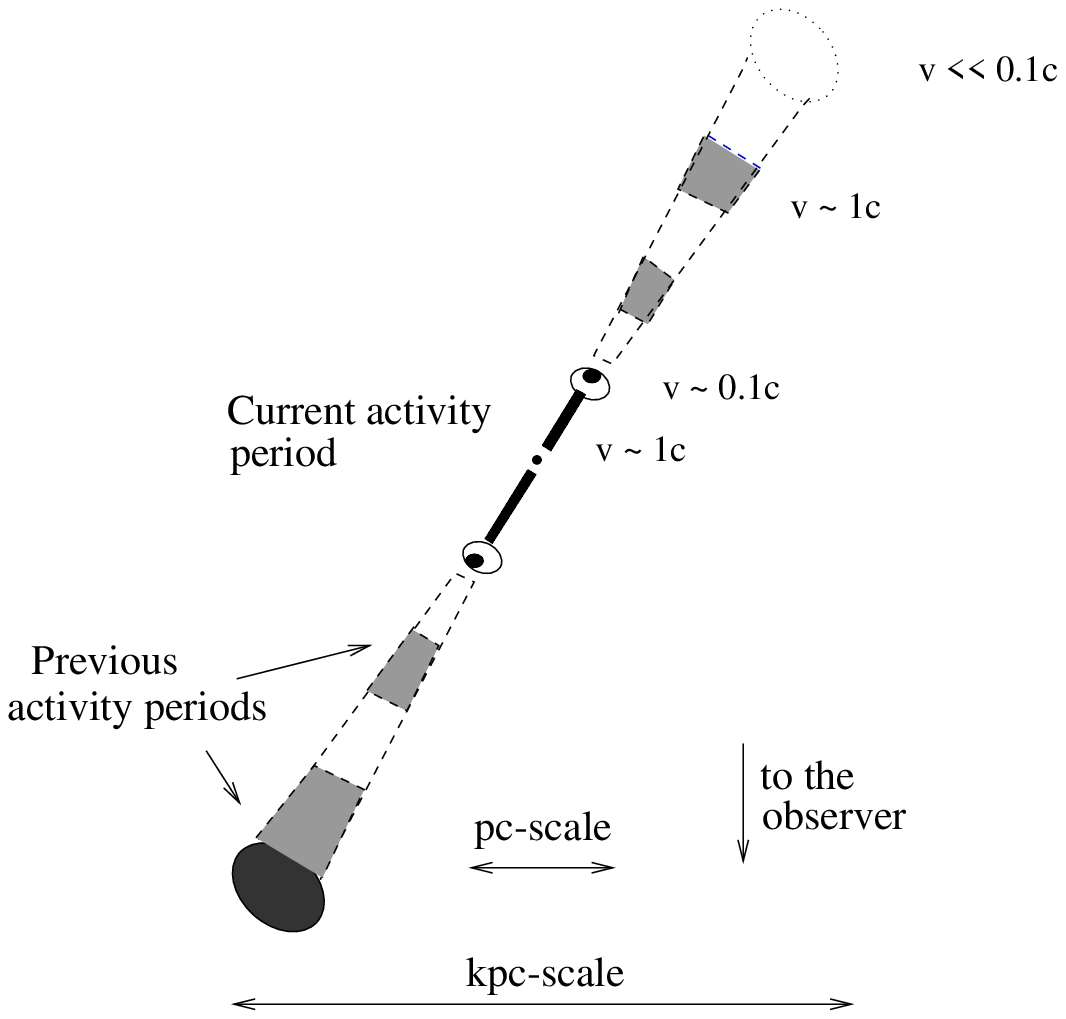}
\caption{{\footnotesize Diagram illustrating a possible explanation of
the  extended emission in 0108+388. {\it  Light grey} areas represent
material emitted from the central engine in previous periods of activity
which propagates through a low density cocoon.
{\it Dark grey} areas represent working surfaces against high density 
circumnuclear and IGM gas from which we get significant radio emission 
on pc- and kpc-scales respectively.}} 
\label{Fig.3}
\end{figure}
%________________________________________________________________

The fact that we see kpc-scale structure only on one side of 0108+388 might
be understood  as a natural  consequence  of  intermittent  activity in the
nucleus  (although Baum et al.  \cite{baum}  discuss other  possibilities).
If the on-off cycle is short then separate  regions of activity (light grey
areas  in  Fig.~\ref{Fig.3})  will  propagate  out  towards  the  kpc-scale
hotspots.  Due to light travel time  effects we are viewing the far side of
the source at an  earlier  time than the near  side.  It may be that we are
viewing the Eastern, near side kpc-scale hotspot at a time when it is being
supplied by jet material and the Western, far side kpc-scale  hotspot while
unsupplied.  Such an  unsupplied  hotspot  fades  very  fast  as  electrons
diffuse  to  regions  of lower  density  and  magnetic  field  and  radiate
inefficiently.  Given the size of the  Eastern  hotspot at the lowest 5 GHz
contour  (Baum et al.  \cite{baum}),  the fact that the Western  hotspot is
not detected at this contour and assuming an electron  backflow velocity of
$0.1c$,  we  calculate  that  the  Eastern  hotspot  must  have  been  left
unsupplied for at least of $2\times  10^{5}h^{-1}$  yrs,  providing a lower
limit on the on-off cycle time.

An approximate  upper limit to the cycle time can be set by considering the
effects of light  travel  time and jet  propagation.  We first note that in
the proposed model (see  Fig.~\ref{Fig.3})  the kpc-scale  hotspots advance
with low average  speed  (v$<<0.1c$)  through the  relatively  dense IGM or
intercluster  gas;  hence  the  observed  projected  distances  to the  two
kpc-scale  hotspots  are expected to be  approximately  the same.  From the
measured  distance of the Eastern  kpc-scale  hotspot from the core and the
estimated  angle to the line of sight (see Sect 3.2), we can calculate that
this hotspot is viewed  $5\times10^{5}h^{-1}$  yrs earlier than the Western
kpc-scale  hotspot.  The  regions  of jet  activity  (light  grey  areas in
Fig.~\ref{Fig.3})  are assumed to propagate  out at high speed  through the
low density  radio cocoon.  Relativistic  velocities  for this material are
consistent  with the detection of a possible weak  kpc-scale jet feature on
the Eastern  side of the source  (Baum et al.  \cite{baum}).  If the on-off
cycle time was long  compared to the light travel time  difference  then we
would most  likely  observe  either both  kpc-scale  hotspots  supplied  or
unsupplied.  To have a high probability of seeing only one hotspot supplied
we require that the lengths of the periods of activity  must be  comparable
or shorter to the light  travel time (see Fig 3); hence we can  estimate an
upper limit on the cycle time of about $10^{6}h^{-1}$yrs.

We conclude that the pc-scale structure in 0108+388 is very young, but that
activity in this source may be  intermittent.  In this case we are probably
viewing 0108+388 just as a new phase of high  radio-efficiency  activity in
the central engine has begun.  Our results  imply that at least some CSOs
are  recurrent   sources.  In  contrast   other  CSOs  (e.g.  0710+439  and
2352+495)  apparently  show no  extended  emission  or  signs of  recurrent
activity, so the CSO population may contain a mixture of  intermittent  and
non-intermittent sources.

\begin{acknowledgements}  

IO acknowledges support from EU grant
(ERBCIPDCT940087), OSO and S.  Batory Foundation. 
AGP acknowledges support from the European Commission's TMR programme,
Access to Large-Scale Facilities, under contract number ERBFMGECT950012
\end{acknowledgements}

\end{document}